%% file: enumrev.tex
\documentstyle[epsf,amsmath,amsfonts,amssymb,amsbsy,cite]{europhys}
\input euromacr

\begin{document}
\euro{XX}{YYY}{ZZZ}{19990} \Date{} \shorttitle{P.~D. Olmsted,
  Complex fluids in shear flow}
\title{Two-state shear diagrams for complex fluids in shear flow} 
\author{P.~D. Olmsted\footnote {{\tt p.d.olmsted@leeds.ac.uk}}}
\institute{Department
  of Physics and Astronomy and Polymer IRC, University of Leeds, 
 Leeds LS2 9JT, United Kingdom} %
\rec{}{} \pacs{ 
\Pacs{83}{20.Hn}{Rheology: Structural and phase changes}
\Pacs{83}{20.Bg}{Rheology: Macroscopic (phenomenological) theories}
\Pacs{47}{20.Hw}{Fluid Dynamics: Morphological instability, phase changes}
}
\maketitle
\begin{abstract}
  The possible ``phase diagrams'' for shear-induced phase transitions
  between two phases are collected. We consider shear-thickening and
  shear-thinning fluids, under conditions of both common strain rate
  and common stress in the two phases, and present the four
  fundamental shear stress vs. strain-rate curves and discuss their
  concentration dependence. We outline how to construct more
  complicated phase diagrams, discuss in which class various
  experimental systems fall, and sketch how to reconstruct the phase
  diagrams from rheological measurements.
\end{abstract}
\section{Introduction}
There is growing interest in the flow behavior of complex fluids,
including worm-like micellar surfactant solutions
\cite{rehage91,berret94b,Capp+97,grand97,callaghan,schmitt95,pine,NoirLapp97b},
lamellar and ``onion'' surfactant systems
\cite{diat,sierro97,Bonn+98}, and liquid crystals
\cite{safinya91,mather97}. In these and other systems the
microstructure is altered by flow such that a bulk transition,
reminiscent of an equilibrium phase transition, can occur.  Signatures
of this transition are kinks, plateaus, or non-monotonic behavior in
the measured ``flow curve'', that is, the relationship between applied
shear stress and mean strain rate; structural changes observable by
X-ray \cite{safinya91,NoirLapp97b}, neutron
\cite{Capp+97,NoirLapp97b}, or light scattering; explicit measurements
of macroscopically discontinuous flow profiles \cite{callaghan}; and
visual confirmation of phase separation and coexistence
\cite{Capp+97,pine}.
  
{\begin{figure}[!b]
 \displaywidth\columnwidth 
 \epsfxsize=3.0truein
 \centerline{\epsfbox[100 40 545 364]{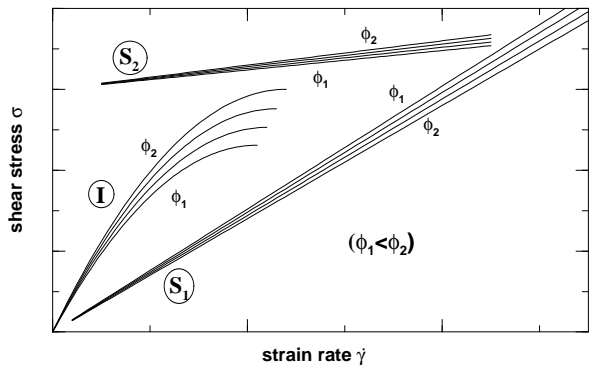}}
 \caption{Steady state constitutive curves $\sigma(\dot{\gamma},\phi)$ for a
   hypothetical fluid with a shear-thinning quiescent state
   \textbf{I}, a shear-induced phase \textbf{S$_1$}, and a
   shear-induced ``gel'' \textbf{S$_2$} with an apparent yield stress.
   All flow branches have concentration dependences as shown, and are
   only stable at the points shown.}
 \label{fig:toys}
  \end{figure}}
Despite the surfeit of experiments, theories have been limited to a
few systems (micelles \cite{wormgel,spenley96} and liquid crystals
\cite{hess76,bruinsma91,olmstedgold,olmstedlu}), and few have
attempted to calculate an entire phase diagram for a complex fluids
solution in flow \cite{olmstedlu}. In a remarkable paper, Schmitt
\emph{et al.}  classified the instabilities that can occur in complex
fluid solutions, and clarified the relationship between the nature of
the instability (either in primarily the concentration, ``spinodal'',
or the strain rate, ``mechanical''), the shape of the flow curves, and
the orientation of the interface that initially develops during the
instability \cite{schmitt95}. However, as noted in \cite{schmitt95},
the orientation of an interface in an initial instability may or may
not be relevant to the orientation of the macroscopic phase separated
state. In this work I outline the different macroscopic phase diagrams
that can occur in complex fluid solutions in planar shear flow and
describe how phase diagrams determine ``flow curves'', the relation
between applied shear stress and measured mean strain rate. No attempt
is made to pose or solve any specific dynamical models (see
\cite{olmstedlu,goveaspine99}), but rather to explore the consequences
of possible phase diagrams and provide a phenomenological framework
within which to understand the rapidly growing body of rheological
experiments.

\section{Phenomenology of Phase Coexistence in Shear Flow}
Calculations of phase behavior requires first determining the relevant
microstructure of the quiescent and flow-induced phases, and deriving
equations of motion for the appropriate variables (momentum, mass
concentration $\phi$, and variables such as liquid crystalline or
crystalline order, or aggregate shape and size distributions).  These
equations of motion determine the steady state flow curves
corresponding to different microstructures. For example,
fig.~\ref{fig:toys} shows a hypothetical shear-thinning fluid
\textbf{I} which can be sheared into either a less viscous fluid
\textbf{S$_1$}, or a gel-like state \textbf{S$_2$}.  For this
particular fluid the shear-induced phases are stabilized only in
finite flow; systems near equilibrium phase transitions such as liquid
crystals may have multiple stable flow curves in the limit of zero
flow \cite{olmstedlu}, while other systems such as dilute wormlike
micelles \cite{pine} and ``onions'' \cite{diat,sierro97,Bonn+98}
apparently have flow branches which are only stabilized in finite
flow.  The task of theory, and indeed of experiments, is to understand
how material on different flow curves may coexist, and what controls
the stability of one phase with respect to another.

At coexistence a fluid partitions into phases with, in principle,
different concentrations.  In the fluid of fig.~\ref{fig:toys} the
\textbf{I} and \textbf{S$_1$} states could coexist at the same shear
stress (which would require the interface between phases to lie in the
velocity-vorticity plane); as could the \textbf{S$_1$} and
\textbf{S$_2$} states; the \textbf{I} and \textbf{S$_1$} phases could
not coexist at a common shear stress.  Phase coexistence among all
three possible pairs of states is conceivable at a common strain rate
(requiring an interface lying in the velocity--velocity-gradient
plane).  Phase diagrams may be constructed by determining the states
on distinct stable flow curves for which the driving force for
particle exchange vanishes, and for which a mechanically stable
stationary interfacial solution between phases can be found. This
procedure has been developed for dynamical models which incorporate
both smooth \cite{olmstedgold,spenley96,olmstedlu} and sharp
\cite{ajdari98,goveaspine99} interfaces. In the former case gradient
terms are included in the equations of motion, while in the latter
case a mechanical condition on interface stability is required.
Complete phase diagrams have been calculated for a model system of
rigid rod suspensions, for both common stress and common strain rate
geometries \cite{olmstedlu}.  Unfortunately, we still lack methods for
choosing from several candidate possibilities for phase coexistence.

To begin, consider a complex fluid solution which possesses two
distinct phases under shear flow: we denote these \textbf{I} and
\textbf{S} (for the example fluid of fig.~\ref{fig:toys} these could
be any of the \textbf{I}, \textbf{S$_1$} and \textbf{S$_2$} states).
We assume this fluid has steady-state macroscopic phase coexistence
observed experimentally and hence has a phase diagram which may, in
principle, be calculated from the relevant dynamical equations of
motion. For simplicity we ignore interesting complications associated
with secondary-flow or dynamical instabilities which can induce
non-stationary oscillating or chaotic steady states; and ignore any
effects of curved geometries.

Given such a fluid, there are two possible geometries for phase
coexistence: (A) common stress phase separation, for which phases
coexist in the flow gradient direction, and (B) common strain rate
phase separation, for which phases coexist in the vorticity direction.
For each case, flow can induce a transition to either a less viscous
phase (A1, B1) or a more viscous phase (A2, B2).  These geometries and
flow curves are shown in fig.~\ref{fig:general}. To understand the
concentration dependence of these flow curves we must examine the
entire phase diagrams: some possibilities are shown in
fig.~\ref{fig:all}. Upon increasing the concentration the \textbf{S}
phase may be stabilized at either weaker flows (A1, B1, A2,
\emph{etc.})  or stronger flows (A1', B1', A2', \emph{etc.}).
{\begin{figure}[!bt] \displaywidth\columnwidth \epsfxsize=5.0truein
    \centerline{{\epsfbox[0 0 351 209]{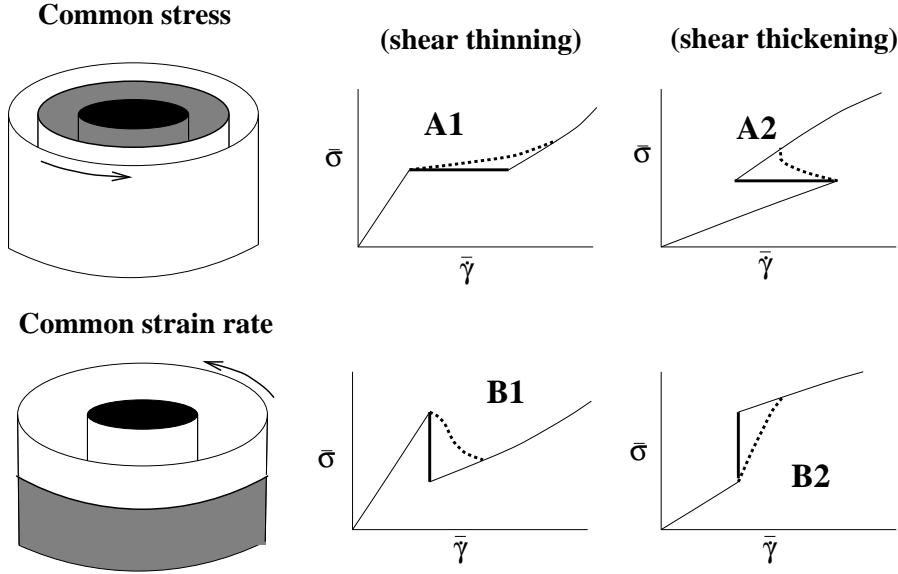}}}
\caption{Fundamental rheological signatures for phase separation at
  common stress (top row, A1 and A2) or common strain rate (bottom
  row, B1 and B2), for essentially shear thinning (middle, A1 and B1)
  or shear thickening (right, A2 and B2) transitions. $\bar{\sigma}$
  and $\bar{\dot{\gamma}}$ are the imposed (measured) shear stress and
  strain rate, respectively. The thick solid lines denote the two
  phase region if $\delta\phi=0$, and the broken lines denote the
  two-phase region in the event of a finite width biphasic region
  $|\delta\phi|> 0$. }
\label{fig:general}
\end{figure}}

{\begin{figure}[!b] 
    \displaywidth\columnwidth 
    \epsfxsize=5.8truein
    \centerline{{\epsfbox[0 0 303 170]{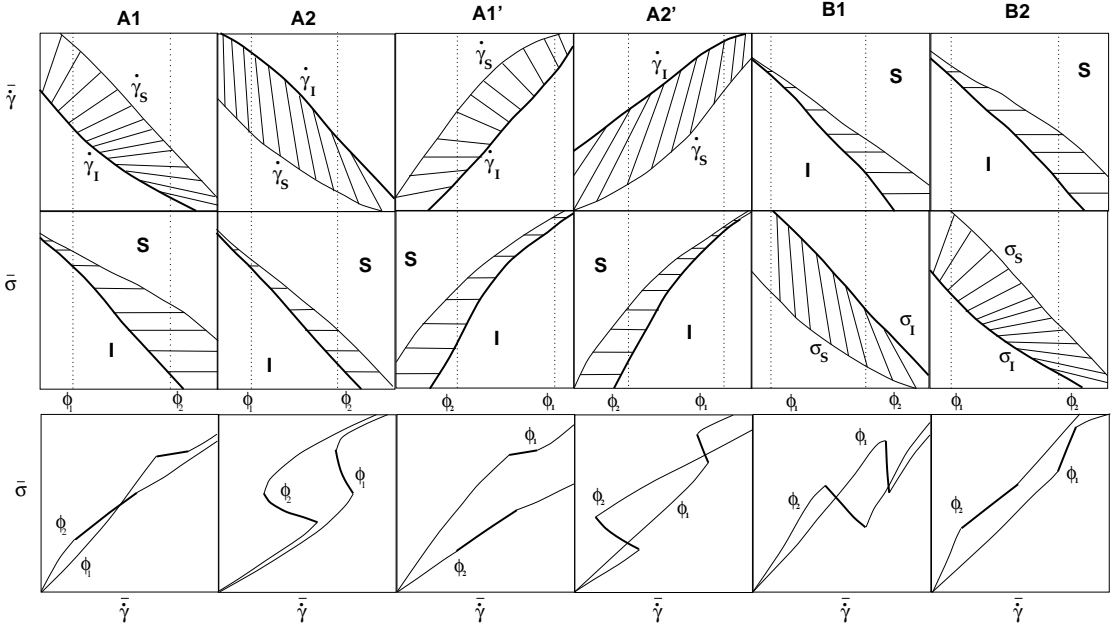}}}
    \caption{``Fundamental''  phase diagrams and flow curves.
      These have not been calculated, but would be calculated from the
      homogeneous flow curves (\emph{e.g.} fig.~\ref{fig:toys}) by the
      methods of \cite{olmstedlu} or \cite{ajdari98,goveaspine99}.
      The phase boundaries $\dot{\gamma}_S, \dot{\gamma}_I, \sigma_S,
      \sigma_I$ in the top two rows are coincident with the kinks in
      the flow curves.  Top row: shear diagrams in the
      $\dot{\gamma}\!-\!\phi$ plane.  Middle row: shear diagrams in
      the $\sigma\!-\!\phi$ plane.  Thick lines in both planes
      correspond to the \textbf{I} phase boundary. Bottom row: flow
      curves $\sigma(\dot{\gamma})$ for representative mean
      concentrations $\phi_1$ and $\phi_2$, with thick lines denoting
      the phase coexistence region; in each case $\phi_1$ has a
      narrower window of coexistence in the relevant control variable
      plane than does $\phi_2$.  Tie lines in the
      $\dot{\gamma}\!-\!\phi$ plane move to higher strain rates with
      increasing stress. \textbf{A1} and \textbf{A2}: phase separation
      at common stress with the \textbf{S} phase stabilized at a lower
      stress for increasing concentration.  \textbf{A1'} and
      \textbf{A2'}: phase separation at a common stress, with
      \textbf{S} phase stabilized at a higher stress for increasing
      concentration.  \textbf{B1} and \textbf{B2} phase separation at
      a common strain rate.  }
    \label{fig:all}
  \end{figure}}

\section{Common Stress Phase Separation}
We first consider coexistence at a common stress (A1, A2, A1', A2'),
for which the mean strain rate $\bar{\dot{\gamma}}$ and mean
concentration $\bar{\phi}$ at stress $\sigma$ are partitioned
according to
\begin{eqnarray}
  \bar{\dot{\gamma}} &=& \alpha\,{\dot{\gamma}}_I(\phi_I) +
  (1-\alpha)\,\dot{\gamma}_S(\phi_S),\label{eq:levsig} \\
  \bar{\phi} &=& \alpha\,{\phi}_I +  (1-\alpha)\,\phi_S,\label{eq:levphi}
\end{eqnarray}
with $\alpha$ the fraction of material in the \textbf{I} phase.  Phase
coexistence occurs in a region in the $\sigma\!-\!\phi$ plane, with
horizontal tie lines connecting the coexisting points
$(\phi_I,\dot{\gamma}_{I})$ and $(\phi_S,\dot{\gamma}_S)$. This may
also be represented in the $\dot{\gamma}\!-\!\phi$ plane.  The lines
$\{\phi_I(\sigma),\phi_S(\sigma), \dot{\gamma}_I(\phi_I(\sigma)),
\dot{\gamma}_S(\phi_S(\sigma))\}$ bound the phase coexistence regions.
The slope of tie lines in the $\dot{\gamma}\!-\!\phi$ plane reflects
the compositions of the two phases; vertical tie lines imply equal
concentrations, while a non-infinite slope implies different
concentrations.  The flow curves $\sigma({\dot{\gamma}},\bar{\phi})$
can be calculated from the phase diagrams and eqs.~\ref{eq:levsig}
and~\ref{eq:levphi} by applying the lever rule.  They are non-analytic
at the stresses and strain rates which bound the biphasic region, and
would be obtained in experiment at a given mean concentration
$\bar{\phi}$.  In the biphasic region the stress increases by crossing
successive tie lines in the $\dot{\gamma}\!-\!\phi$ plane, and varies
according to the tie line spacing and splay \cite{olmstedlu}.  From
eqs.~\ref{eq:levsig} and~\ref{eq:levphi} one can calculate the shape
of the stress ``plateau'' \cite{schmitt95}:
\begin{equation}
  \left.\frac{d\sigma}{d{\bar{\dot{\gamma}}}}\right|_{\bar{\phi}} = \left[
    \frac{\alpha}{\eta_I} + \frac{1-\alpha}{\eta_S} -
    m(\sigma,\bar{\phi})\left\{\frac{1-\alpha}{\dot{\gamma}'_S\eta_S} +
      \frac{\alpha}{\dot{\gamma}'_I\eta_I}\right\} 
\right]^{-1},
\end{equation}
where $m(\sigma,\bar{\phi})$ is the slope of the tie line,
$\eta_{k}=\partial\sigma/\partial{\dot{\gamma}}$ is the local
viscosity of the $k$th branch, and
$\dot{\gamma}'_{k}=\partial\dot{\gamma}_{k}/\partial\phi$.
Experiments on worm-like micelles have revealed that some systems can
be ``superstressed'' to a metastable state, with the stress
entering the biphasic region \cite{grand97,berret94b,Berr97}.

For an {A1} (or {A1'}) shear-thinning transition the strain rate
increases as the stress is increased through the two phase region. The
stress through the two-phase region increases less for mean
concentrations $\bar{\phi}$ for which the biphasic window is narrow
(\emph{e.g.} $\phi_1$) or the tie lines are steep in the
$\dot{\gamma}\!-\!\phi$ planel, and deviates more from constant for a
wider biphasic region (\emph{e.g.} $\bar{\phi}=\phi_2$) with more
vertical tie lines. Hence, a signature of substantial concentration
difference $\delta\phi\equiv\phi_I-\phi_S$ between coexisting phases
is a large increase in stress through the biphasic region.  This is
consistent with calculations on a model for rigid rods in shear flow
\cite{olmstedlu}, and measurements on shear-thinning wormlike
micelles
\cite{rehage91,berret94b,Capp+97,grand97,callaghan,schmitt95}.
In dilute micelles $\delta\phi$ is typically negligible and the stress
plateau is nearly flat; while $\delta\phi$ increases for concentrated
solutions near an underlying isotropic-nematic transition and the
stress plateau acquires a shape.

For the {A1} fluid the \textbf{S} phase can have a larger
($m(\sigma,\bar{\phi})>0$) or smaller ($m(\sigma,\bar{\phi})<0$)
strain rate than the \textbf{I} phase. In the latter case the
\textbf{S} phase has a larger effective viscosity, and for a small
enough $\delta\phi$ the fluid crosses over to a characteristic
shear-thickening fluid, {A2}. In this case the strain rate actually
decreases as the stress is increased through the two phase region.
That is, the system remains in the \textbf{I} phase below a critical
strain rate at which a band of more viscous material \textbf{S}
develops whose small strain rate \emph{reduces} the measured strain
rate. Hence the system enters the biphasic region in the
$\dot{\gamma}\!-\!\phi$ plane at the high strain rate, and traverses
from top to bottom.  Upon increasing the stress the system converts
into the \textbf{S} phase with, in general, changing \textbf{I} and
\textbf{S} concentrations, until the \textbf{I} phase disappears at
the lower boundary of the two-phase envelope in the
$\dot{\gamma}\!-\!\phi$ plane. For higher stresses the system follows
the constitutive branch for the \textbf{S} phase and traces a vertical
path in the $\dot{\gamma}\!-\!\phi$ plane.  Hence the flow curve
$\sigma({\dot{\gamma}})$ has a distinct \textsf{S} shape.  As with
flow {A1}, the shape of the flow curve in the biphasic region depends
on the slope of the tie lines (\emph{i.e.}  $\delta\phi$) in the
$\dot{\gamma}\!-\!\phi$ plane: vertical tie lines $\delta\phi=0$ imply
a vertical jump in $\sigma(\dot{\gamma})$, while finite $\delta\phi$
and flatter tie lines imply a gentler slope for
$\sigma({\dot{\gamma}})$.  Flow {A2'} has been seen in
shear-thickening wormlike micelles \cite{pine}, and calculated using a
phenomenological model \cite{goveaspine99}.  Flow {A2} (or {A2'}) has
been seen in surfactant onion crystals under shear, although the band
geometry has not been verified \cite{sierro97}.

We have drawn (A1,A2,A1',A2') with finite biphasic regions at zero
stress, corresponding to a perturbation of an existing phase
transition. However, fluids such as the example fluid in
fig.~\ref{fig:toys} would have phase coexistence only above a finite
stress.  The construction of these phase diagrams implies that tie
lines move to higher strain rates with increasing stress
(``conventional''). Although intuitive, the reverse
(``unconventional'') does not violate any physical laws.  For example,
a version of A2' with unconventional positive-sloped tie lines, would
imply a $\sigma\!-\!\phi$ biphasic region moving to smaller $\phi$
with increasing $\sigma$.  However, a version of A2' with
negative-sloped tie lines yields a nonsensical phase diagram.

\section{Common Strain Rate Phase Separation}
For coexistence at a common strain rate the shear stress is
partitioned according to
\begin{equation}
  \bar{\sigma} = \alpha\,{\sigma}_I(\phi_I) +
  (1-\alpha)\,\sigma_S(\phi_S), \label{eq:levgam}
\end{equation}
and the two main classes of flows are shown as B1 and B2. We do not
show analogous diagrams B1' and B2', for which the coexistence strain
rate increases with increasing concentration. For flow B2 the stress
increases through the two phase region and the transition is to either
a thicker or (for large enough $\delta\phi$, as with flow {A1})
thinner phase.  The shape of the stress through the two phase region
is given, from eqs.~\ref{eq:levphi} and~\ref{eq:levgam}
\begin{equation}
  \left.\frac{d\bar{\sigma}}{d{\dot{\gamma}}}\right|_{\bar{\phi}} = 
    \eta_I\,\alpha + \eta_S\,(1-\alpha) -
    m(\dot{\gamma},\bar{\phi})\left\{\frac{\eta_S(1-\alpha)}{\sigma'_S} +
      \frac{\eta_I\,\alpha}{\sigma'_I}\right\},
\end{equation}
where $m(\dot{\gamma},\bar{\phi})$ is the slope of the tie line and
$\sigma'_{k}=\partial\sigma_{k}/\partial\phi$.  In the limit of no
concentration difference ($\delta\phi=0$ or
$m(\dot{\gamma},\bar{\phi})=\infty$), $\sigma(\dot{\gamma})$ is
vertical through the two phase region. Flow B1 is a shear thinning
transition, and the stress decreases through the two-phase region.
Banding geometries and flow curves similar to B1 were reported in an
onion system \cite{Bonn+98}, although this may not have been steady
state.  B2 has been reported in an onion system \cite{diat}.

\section{Applications}
The shear diagrams presented do not exhaust the possibilities; these
flow behaviors can be combined smoothly in many ways.  We present two
possibilities in fig.~\ref{fig:comb}.  Fluid C1 phase separates at a
common stress, and $\delta\phi$ widens with increasing concentration,
crossing over from shear thickening (A2) to shear thinning (A1) as the
concentration increases.  Fluid C2 phase separates at a common strain
rate, with a coexistence width $\delta\phi$ that narrows with
increasing $\phi$, combining B1' and B2' behavior.

{\begin{figure}[!b ] \displaywidth\columnwidth \epsfxsize=4.5truein
    \centerline{\epsfbox[0 0 308 191]{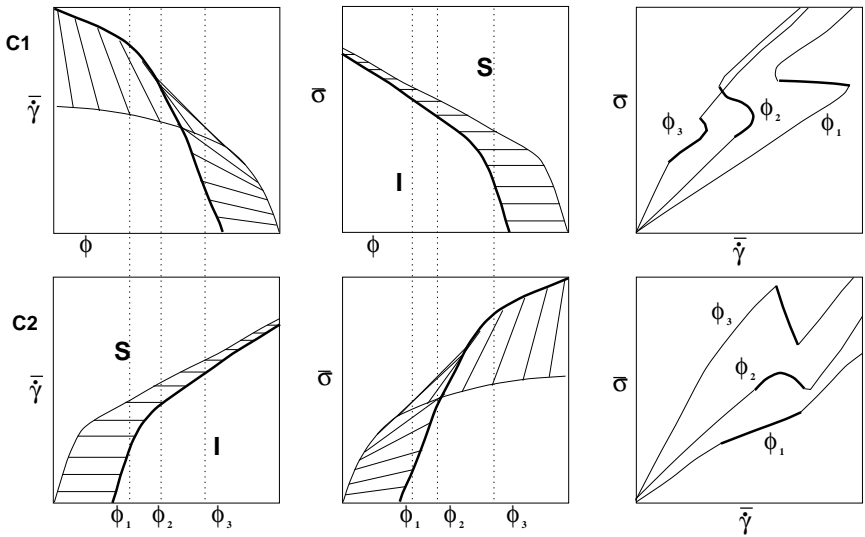}}
\caption{More complex possibilities for phase separation at common
  stress (C1: top row) or common strain rate (C2: bottom row).
  $\phi_1, \phi_2$, and $\phi_3$ are representative mean
  concentrations.}
\label{fig:comb}
\end{figure}}
Typical experiments extract a series of steady state flow curves for
different concentrations $\{\phi_i\}$. Consider coexistence at common
stress.  The flow curves have kinks at the boundaries of the biphasic
region, $\{(\sigma_{Ii},\dot{\gamma}_{Ii},\phi_i)\}$ and
$\{(\sigma_{Si},\dot{\gamma}_{Si},\phi_i)\}$, which should be
corroborated optically or otherwise.  The strain rates and
concentrations of the coexisting states may be determined as follows.
One first fits curves $\sigma_I(\phi),\sigma_S(\phi),
\dot{\gamma}_I(\phi),\dot{\gamma}_S(\phi)$ to the values extracted
from the kinks.  A horizontal tie line connecting $(\phi_I,\phi_S)$ at
a given stress $\sigma_{\ast}$ may then be read off the biphasic
boundaries,
$\sigma_I(\phi_{I\ast})=\sigma_S(\phi_{S\ast})=\sigma_{\ast}$ in the
$\sigma\!-\!\phi$ plane; and the corresponding tie lines on the
$\dot{\gamma}\!-\!\phi$ plane connects the points
$\dot{\gamma}_I(\phi_{I\ast})$ and $\dot{\gamma}_S(\phi_{S\ast})$. In
this way the complete shear diagrams and characteristics of coexisting
states, may be constructed. As a check, the shape of
$\sigma({\dot{\gamma}})$ through the two-phase region may be computed
by traversing the biphasic regions of the $\dot{\gamma}\!-\!\phi$ and
$\sigma\!-\!\phi$ planes and using the lever rule to construct the
mean strain rate via eq.~(\ref{eq:levsig}). This analysis is necessary
if the coexisting concentrations cannot be determined directly and, in
the two phase region, the flow curve has a slope appreciably different
from zero (common stress phase separation) or infinity (common strain
rate phase separation).
  
Returning to the hypothetical fluid of fig.~\ref{fig:toys}, we expect
common stress phase separations \textbf{I-S$_1$} and
\textbf{S$_1$-S$_2$} to be classes A1 and A2 (or A1' and A2'),
respectively, with the former disappearing at higher stresses; and
common strain rate phase separations \textbf{I-S$_1$},
\textbf{I-S$_2$}, and \textbf{S$_1$-S$_2$} to be classes B1, B2, and
B2 (or B1', B2', B2'), respectively.  Since flow is necessary to
stabilize the \textbf{S$_1$} and \textbf{S$_2$} phases, the biphasic
window in this system would appear at a finite stress or strain rate.
Note that the accessibility and even stability of composite flow
curves depends on the control variable of the rheometer. For example,
composite flow curves with negative slope
$d{\sigma}/d{\dot{\gamma}}<0$ (\emph{e.g.} A2 and B1) could be
mechanically unstable \cite{schmitt95}, although ref.~\cite{pine}
accessed such a curve under controlled stress conditions. Finally,
kinetic possibilities such as metastability and hysteresis are sure to
enrich the relatively simple scenarios proposed above.

\stars It is a pleasure to thank David Lu, Armand Ajdari, Ovidiu
Radulescu, Jacqueline Goveas, and Tom McLeish for encouragement and
enjoyable discussions.

\end{document}

%% file: euromacr.tex
\def\stars{\bigskip\centerline{***}\medskip}

\newif\ifboo \boofalse
